\documentclass[%
 groupedaddress,
 nofootinbib,
 amsmath,amssymb,
 aps,
 pra,
 twocolumn,
floatfix,
10pt
]{revtex4-2}

\usepackage{hyperref}
\usepackage{graphicx}% Include figure files
\usepackage{dcolumn}% Align table columns on decimal point
\usepackage{bm}% bold math
\begin{document}

\title{Real-time decoding of quantum error correction codes using high-performance computing}

\newcommand{\nudt}{{College of Computer Science and Technology, National University of Defense Technology, Changsha 410073, China}}

\author{Lingling Lao}\affiliation{\nudt}
\author{Qiang Wang} \affiliation{\nudt}
\author{Yuanqi Liu} \affiliation{\nudt}
\author{Yantong Liu} \affiliation{\nudt}
\author{Haowen Wang} \affiliation{\nudt}
\author{Yitao Chen} \affiliation{\nudt}
\author{Yankang Zhao} \affiliation{\nudt}
\author{Zhenwei Wu} \affiliation{\nudt}
\author{Wei Zhang} \affiliation{\nudt}
\author{Yong Dong} \affiliation{\nudt}
\author{Yingwen Liu} \affiliation{\nudt}
\author{Mingche Lai} 
\email{mingchelai@nudt.edu.cn}\affiliation{\nudt}
\author{Junjie Wu} \email{junjiewu@nudt.edu.cn} \affiliation{\nudt}

\date{\today}% It is always \today, today,
             %  but any date may be explicitly specified

\begin{abstract}
Quantum error correction (QEC) is indispensable for building scalable fault-tolerant quantum computers.
Effective QEC demands stringent real-time decoding: the decoder must process syndrome measurements and determine corrections within a time scale—typically on the order of microseconds, to avoid data backlog. 
Scaling to large number of logical qubits further necessitates significant computational resources.
In this work, we propose an architecture, called \emph{THQLink}, for real-time decoding of quantum error correction codes using high-performance computing (HPC) resources.
The network connecting the HPC and the control system of quantum processing unit (QPU) is built on TH-Express and can be adapted to different quantum technologies and their associated control stacks. 
We report a round-trip latency of 2.944~$\mu$s on average, with an incremental overhead of 130~ns per additional hop. 
Using a parallel window strategy, we demonstrate real-time decoding (1~$\mu$s per QEC round) of the surface code up to distance 19 using a matching-based decoder on CPUs.
Our work presents a scalable framework for real-time decoding in fault-tolerant quantum computing. 
It can be readily applied to quantum-centric supercomputers that feature tight integration between QPU and HPC resources, thereby enabling efficient support for hybrid quantum-classical algorithms and computation-intensive workloads offloaded from the QPU.
\end{abstract}

\maketitle

%\tableofcontents

\section{INTRODUCTION}
\label{sec:introduction}

Quantum computing holds the promise of solving problems intractable for classical supercomputers, yet its practical realization hinges critically on the ability to suppress physical errors through quantum error correction (QEC)  ~\cite{shor1995scheme,steane1996error,terhal2015quantum}. 
The key idea of QEC is to protect information by redundantly encoding many physical qubits into one logical qubit and periodically measure auxiliary qubits to extract error syndromes without collapsing the logical state~\cite{gottesman1997stabilizer,terhal2015quantum}. 
A classical decoder processes the syndrome data to diagnose the most probable errors and prescribe corrective actions~\cite{dennis2002topological}. 
To be viable for large-scale quantum computing, a decoder must satisfy three stringent and often conflicting requirements. 
First, it should achieve accuracy approaching the optimal decoder, thereby minimizing the qubit overhead needed to reach a target logical error rate~\cite{fowler2012surface}. 
Second, it must process syndromes within the hardware's generation clock rate, typically around 1~$\mu$s for superconducting qubits, to prevent an exponential data backlog \cite{terhal2015quantum,skoric2023parallel}. 
This processing time includes both the actual decoding computation and the communication latency between the decoder and the quantum system controller(QSC). 
Meeting the real-time budget necessitates that the computing unit hosting the decoder be either tightly coupled with or physically integrated into QSC~\cite{google2025quantum,caune2026demonstrating}.
Third, the decoder should be scalable to support hundreds or thousands of logical qubits at large code distances for utility-scale quantum computing.

Various decoding algorithms have been developed in both software and dedicated hardware to satisfy one or all the requirements. 
Minimum‑weight perfect matching (MWPM) is a widely used decoder for topological codes \cite{dennis2002,higgott2022pymatching}. 
Sparse blossom is a highly optimized sequential implementation of MWPM designed for fast single-core performance, it decodes a distance-17 surface code with 1 us per QEC round~\cite{higgott2025sparse}. 
Fusion blossom is a parallel MWPM solver to leverage multi-core CPUs, enabling it to decode surface codes up to distance 33 with sub-microsecond per-round latency on a 64-core processor \cite{wu2023fusion}.
The Union‑Find decoder achieves near‑linear time complexity but typically sacrifices some accuracy compared to MWPM \cite{delfosse2021almost}. 
Belief propagation with ordered statistics decoding (BP+OSD) 
is applicable to various codes such as surface codes and bicycle codes \cite{Poulin2008iterative}. Nevertheless, the post‑processing step of OSD may become a bottleneck for real‑time performance. 
Relay‑BP introduces asymmetric memory strengths to enable faster and more accurate convergence to the final error belief \cite{Muller2025-tt}. It achieves roughly a 10‑fold improvement in accuracy over BP+OSD, while maintaining or even improving the speed of standard BP.

Except the conventional algorithms, machine learning (ML) decoding is emerging as a highly viable solution because it
can learn directly from data without needing to fit a precise noise model, allowing them to adapt to the complex and unpredictable noise present in real hardware and thereby achieve near-optimal logical error rates.
Decoders based on reinforcement learning \cite{olle2024simultaneous, sweke2021reinforcement, matekole2022decoding, he2025discovering}, graph neural networks \cite{lange2025datadriven, gong2024graph, maan2025machine}, diffusion models \cite{liu2025decoding, cao2025generative}, and transformers \cite{bausch2024learning, senior2025scalable,zhang2026learning,choukroun2022error, cohen2025hybrid} are being explored.

These software decoders often run on CPUs or GPUs and face prohibitive and non-deterministic delays when connecting to QSC via TCP/IP Network. 
While Peripheral Component Interconnect Express (PCIe) is widely adopted for high-speed peripheral interconnects, its chassis-confined electrical reach creates a fundamental roadblock to cross-node scaling of QEC decoding resources.

On the other hand, decoders implemented on dedicated classical hardware—such as field-programmable gate arrays (FPGAs) \cite{barber2025real,maurya2025fpga, maurer2025real,ziad2025local,wu2025micro,das2020scalable}or application-specific integrated circuits (ASICs) \cite{barber2025real} can be naturally integrated with the control system  of quantum processors and offer superior decoding speed to meet the real-time latency requirements.
FPGAs are widely used for their reconfigurability and ability to achieve high throughput. 
The collision clustering decoder on an FPGA achieves megahertz decoding speeds for surface codes up to distance 21 \cite{barber2025real}.
For BP+OSD, a parallel architecture on a single FPGA can operate at 200 MHz for surface codes up to distance 9 \cite{maurya2025fpga}.
% IBM has successfully demonstrated running its 
The Relay-BP algorithm on an AMD FPGA, and the GARI decoder for correlated errors has shown a preliminary FPGA implementation with an average latency of just 273~ns \cite{maurer2025real}.
ASICs are fixed-function chips that provide the higher performance and lower energy consumption than FPGAs.
The ASIC implementation of the collision clustering decoder decodes a distance‑23 surface code in 0.24~$\mu$s, occupying just 0.064 mm$^2$ of die area and consuming only 7.72 mW of power~\cite{barber2025real}.
While impressive in speed and energy efficiency, this hardware-firmware co-design is inherently rigid. Any modification to the QEC code, noise model, or decoding algorithm necessitates costly and time-consuming chip redesigns.
Google's AlphaQubit 2 demonstrated that transformer-based decoders can achieve near-optimal logical error rates and enable real-time decoding for surface codes up to distance 11 on commercial TPUs \cite{senior2025scalable}. 
However, its reliance on specialized machine-learning accelerators creates a closed ecosystem that is difficult to adapt, or integrate into standard quantum computing  workflows.
To summarize, custom hardware such as FPGAs, ASICs, and TPUs delivers peak deterministic performance but sacrifices scalability or programmability or adaptability. 

NVIDIA's recently NVQLink architecture offers a foundational blueprint for tightly coupling QPUs with HPC resources via a low-latency, deterministic interconnect reporting a mean round-trip latency of 3.839~$\mu$s over RoCE Ethernet\cite{caldwell2025platform}, and balancing performance with scalability, programmability, and support for heterogeneous QPU modalities. 
However, its design prioritizes commercial deployability: it retains generic Ethernet protocol overhead and current validation is limited to pure network loopback tests without real-time decoding demonstration.

In this work, we introduce a scalable real-time QEC decoding architecture, \emph{THQLink}, that leverages a general-purpose HPC cluster as its primary computational engine, tightly coupled to the QSC via a high-speed interconnection network, TH-Express. 
Our interconnect strips redundant protocol fields, achieving a measured mean round-trip latency of 2.944~$\mu$s --- a $23.3\%$ reduction over NVQLink. Critically, THQLink supports seamless scaling to additional decoding compute nodes via its switching fabric, with a measured incremental latency overhead of just 130~ns on average per additional network hop. For the first time, we demonstrate a full closed-loop decoding pipeline spanning FPGA-driven transmission of syndrome data, low-latency transport, decoding, and correction feedback to the FPGA. With a carefully designed parallel window decoding strategy and optimized software implementations, 
We demonstrate real-time (1~$\mu$s per round) decoding requirement at scale using HPC while offering flexibility and scalability that are challenging to achieve with hardware-fixed alternatives.

\section{Background}
\label{sec:background}

\subsection{Quantum error correction}

Quantum systems are inherently fragile: interactions with the environment cause decoherence, while imperfect quantum operations introduce errors that accumulate throughout a computation. Unlike classical information, quantum states cannot be copied due to the no-cloning theorem, making straightforward classical error correction inapplicable~\cite{nielsen2010quantum}. Quantum error correction (QEC) overcomes this fundamental limitation by encoding a logical qubit into an entangled state of many physical qubits, allowing errors to be detected and corrected without directly measuring the encoded quantum information~\cite{shor1995scheme,shor1996fault,calderbank1996good,terhal2015quantum}.

Among the various QEC architectures proposed to date, topological codes, represented by the surface code~\cite{fowler2012surface,kitaev2003fault,dennis2002topological,bravyi1998quantum} and the color code~\cite{bombin2006topological,kubica2015universal,kubica2018abcs,bombin2015gauge}, have become the predominant choice for near-term QEC demonstrations owing to their local connectivity and compatibility with current hardware. In parallel, emerging quantum low-density parity-check (QLDPC) codes have drawn increasing attention for their potential to achieve higher encoding rates and reduce the physical qubit overhead required for fault tolerance, offering a promising pathway toward large-scale quantum computation~\cite{bravyi2024high,zhao2026towards}. Recent experimental progress has further demonstrated the practical viability of QEC: logical qubits have been operated below the surface-code threshold in both superconducting~\cite{google2025quantum} and neutral-atom~\cite{bluvstein2026fault} platforms, while trapped-ion systems have demonstrated break-even QEC performance~\cite{paetznick2026improved}.

Fault-tolerant quantum computing extends the principles of QEC from protecting quantum memories to executing reliable logical computations. In this scenario, the decoder is no longer merely responsible for identifying errors; it must also provide timely feedback to support logical operations. If decoding cannot keep pace with syndrome generation, syndrome data accumulate into a decoding backlog, potentially stalling the computation and compromising fault-tolerance. Moreover, unlike quantum-memory experiments, where Pauli corrections can generally be deferred through Pauli frame tracking, the execution of non-Clifford gates such as the logical $T$ gate requires the pending Pauli frame to be resolved and the corresponding corrections to be applied before the gate can proceed. These requirements place stringent demands on decoding latency, making low-latency decoding a key enabling technology for scalable fault-tolerant quantum computing.

\subsection{Decoding surface codes}
\begin{figure}[htb!]
    \centering
    \includegraphics[width=1\linewidth]{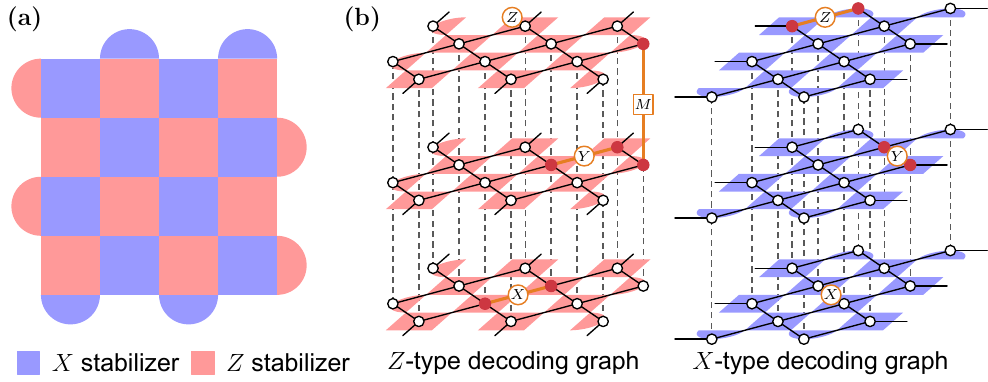}
    \caption{The rotated surface code and its decoding graphs. (a) A distance-5 rotated surface code. (b) Corresponding $Z$- and $X$-type space-time decoding graphs, where horizontal edges represent data-qubit errors and vertical dashed edges represent measurement ($M$) errors. Red nodes denote detection events, and orange edges indicate their matching.}
    \label{fig:sc}
\end{figure}

The surface code encodes a logical qubit into a two-dimensional lattice of physical qubits \cite{fowler2012surface}, as shown in Fig.~\ref{fig:sc}(a). The lattice contains data qubits which store the logical information, and ancilla qubits which perform stabilizer measurements—either 
X-type or Z-type plaquette checks on their four neighboring data qubits. By periodically measuring these stabilizers, the quantum control system extracts syndrome data: a binary string indicating which stabilizer measurements have flipped relative to their expected outcomes. 
A single physical error on a data qubit flips the two adjacent stabilizers, producing a pair of syndrome defects in the spatial lattice. 
Measurement errors add a critical complication: a faulty readout of a stabilizer can produce a false syndrome defect in a single time step, even when no physical error occurred. 
This creates a temporal dimension to the decoding problem, where defects may appear and disappear over successive measurement rounds. 
To distinguish data errors from measurement faults, the decoder must track syndrome changes across multiple rounds, treating a measurement error as a defect that persists for only one time step before vanishing in the next round. 

One can formulate the decoding problem as a graph-theoretic matching problem on the space-time graph shown in Fig.~\ref{fig:sc}(b), where vertices represent syndrome events at specific measurement rounds and edges represent either data errors (connecting defects across space within the same round) or measurement errors (connecting a defect in one round to its absence in the next round via a temporal edge). A widely used decoder based on this formulation is the minimum-weight perfect matching (MWPM) decoder, which seeks a perfect matching that pairs all syndrome defects across the entire space-time graph, yielding the most probable combination of data and measurement errors consistent with the observed syndrome stream. 
The decoder must process each new round of syndrome data in real time—typically within $\sim$1 us for superconducting qubits—to prevent a decoding backlog that would compromise fault tolerance.
This stringent timing constraint has catalyzed extensive research into hardware-accelerated decoders~\cite{barber2025real,ziad2025local}, transformer-based approaches~\cite{ senior2025scalable}, and parallel-window decoding architectures ~\cite{skoric2023parallel, tan2023scalable} to keep pace with kilohertz to megahertz syndrome rates anticipated in near-term quantum devices.

\section{THQLink}
\label{sec:thqlink}
\subsection{Architecture overview}

\begin{figure*}[tbh!]
    \centering
    \includegraphics[width=0.7\textwidth]{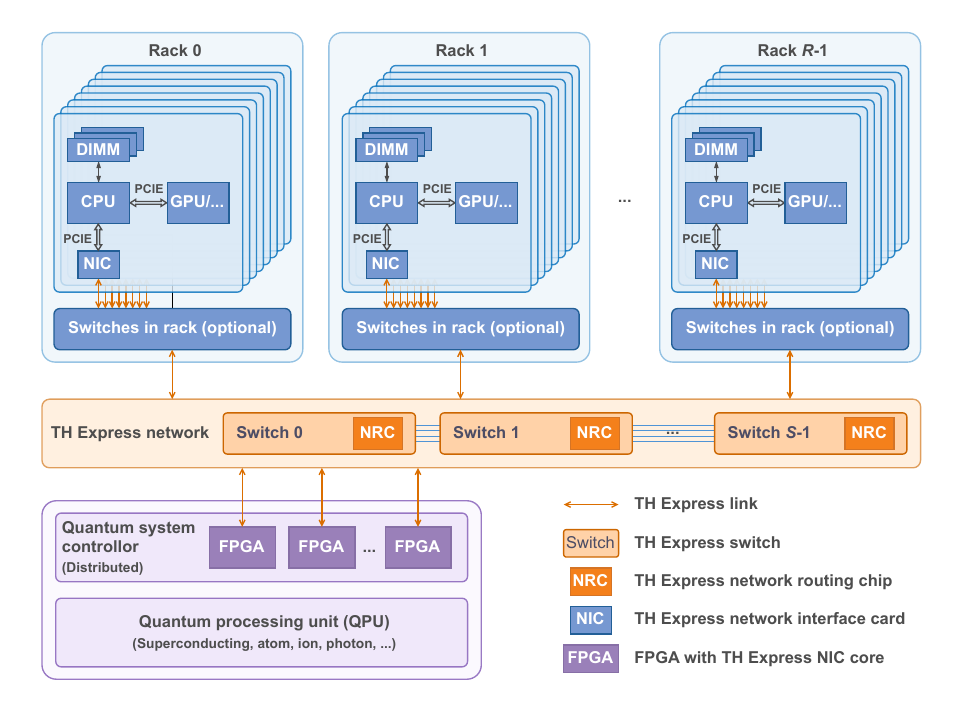}
    \caption{Overview of the \emph{THQLink} architecture.}
    \label{fig:thqlink}
\end{figure*}

The proposed architecture, \emph{THQLink}, integrates a quantum control system with high-performance computing resources to support low-latency, real-time decoding of quantum error-correction codes. 
During quantum execution, the control system performs qubit manipulation and measurement while continuously collecting syndrome information from the quantum processor. 
The resulting syndrome stream is transferred through a high-speed \emph{TH-Express} interconnect to the decoding infrastructure \cite{xu2020scalable, xie2011tianhe}. 
A hierarchy of TH-Express switches connects the quantum control system to multiple HPC racks, providing high-bandwidth and low-latency communication between the syndrome data source and distributed classical computing resources. 
This switched fabric allows decoding traffic to be routed to suitable compute nodes while supporting scalable deployment across multiple racks.

Each HPC rack contains CPU- and/or GPU-based compute nodes that cooperate to execute the decoding pipeline. 
The computing nodes are responsible for both executing the decoding algorithms and handling a range of pre-/post-processing tasks, including communication, syndrome preprocessing, task scheduling, decoder coordination, and the delivery of recovery decisions.
Incoming syndrome data can be partitioned across racks according to logical qubits, code patches, or decoding windows, enabling multiple decoding tasks to execute concurrently. 
After computation, the inferred error corrections are returned through the TH-Express network to the quantum control system. 
By combining scalable high-speed network with high-performance computing, the \emph{THQLink} architecture could increase decoding throughput while preserving the low response latency required for fault-tolerant quantum computing.

\emph{THQLink} is also inherently well suited for the paradigm of quantum-centric supercomputing, in which QPUs and classical HPC resources operate cooperatively within a tightly integrated system. These tasks include hybrid quantum-classical algorithms, error mitigation, calibration, etc.
Consequently, the design not only fulfills the stringent latency and throughput requirements of real-time decoding but also provides a scalable and resilient communication and execution backbone for large-scale hybrid quantum-classical computing platforms.

\subsection{High-speed interconnect}
\label{sec:th-express}

\begin{figure*}[tbh!]
    \centering
    \includegraphics[width=0.8\linewidth]{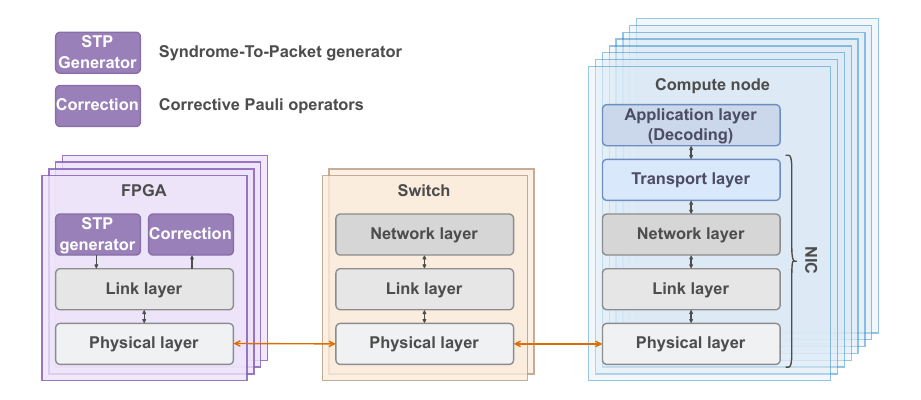}
    \caption{The real-time decoding flow via the high-speed network \emph{TH-Express}.}
    \label{fig:fpga}
\end{figure*}

The high-speed interconnection network is bulit on \emph{TH-Express} developed for Tianhe supercomputers~\cite{xu2020scalable, xie2011tianhe}. 
It is designed to provide a low-latency, high-bandwidth, and scalable communication substrate. 
As shown in Figure \ref{fig:fpga}, the THQLink via TH-Express network is mainly composed of three types of dedicated components: 
network interface cards (NICs) which connect compute nodes to the interconnection network, FPGAs with NIC core which send syndromes and receive corrections, and switches which perform packet switching and routing inside the network.

The NIC implements the host-side protocol functions of TH-Express, covering the physical, link, network, and transport layers, together with the upper-layer communication interfaces exposed to applications and runtime systems. It supports both Mini-Packet (MP), a lightweight short-message communication mechanism, and Remote Direct Memory Access (RDMA) for efficient bulk data movement. 
With up to 800 Gbps of aggregate network bandwidth, sub-0.8~$\mu$s communication latency, RDMA read/write operations, and hardware-offloaded collective communication, the NIC is optimized for latency-sensitive and communication-intensive workloads. 
The FPGAs in QSC implement the physical and link layers of TH-Express together with custom logic, syndrome-to-packet generator and correction for QEC. 

The switch implements the physical, link, and network layers of TH-Express and provides the core switching and routing capability of the interconnection fabric. 
It adopts a non-blocking full-line-rate switching architecture and enables TH-Express to support flexible topologies while maintaining high throughput under dynamic traffic conditions.

% At the network layer, the Switch supports deterministic routing, multipath routing, adaptive routing, wormhole switching, and virtual cut-through switching. At the link layer, it provides link-level retransmission, lightweight reliable transmission, flow control, fault self-healing, and fault-adaptive routing. These mechanisms enable TH-Express to support flexible topologies while maintaining high throughput under dynamic traffic conditions.

Compared with conventional PCIe- or Ethernet-based communication solutions, TH-Express provides stronger topological flexibility and higher protocol efficiency for tightly coupled computing systems. Its dedicated protocol stack reduces software overhead, supports zero-copy data movement, enables hardware-based reliable delivery and traffic control, and allows communication operations to be offloaded from host processors. Therefore, TH-Express is well suited for tight coupling between HPC and QPUs.

\subsection{Real-time decoding via THQLink}
\label{sec:decoding_implementation}

Figure~\ref{fig:fpga} illustrates the closed-loop real-time decoding flow in THQLink. 
%In the proposed architecture, TH-Express provides a low-latency interconnect between FPGAs and compute nodes. 
On the FPGA side, the Syndrome-to-Packet generator encapsulates streaming syndromes into Mini-Packets and transmits them over the lightweight physical/link-layer interface. The packets are then forwarded by the switch. At the host side, the NIC interface passes the syndrome packets through the network and transport them to the application-layer (i.e., the decoder) where the corresponding correction actions are computed.

Before each run, the host constructs the decoding graphs and edge weights, generates lookup tables for mapping recovery edges to correction records, initializes a persistent worker pool and a decoder instance for each window if parallel-window is used, and pre-allocates the required working buffers. These setups remain resident and are reused throughout a run.
As the syndromes arrive, the host prepares the corresponding input blocks of the decoder and stores them in pre-allocated buffers. 
We will evaluate both the global decoding strategy (that decodes all syndromes at once) and the parallel-window decoding strategy ~\cite{skoric2023parallel, tan2023scalable} (that decodes overlapping syndrome rounds to exploit parallelism, more details can be found in Appendix \ref{app:parallel}).
For parallel-window decoding, an A-window task is dispatched as soon as all data required by its core and buffer regions are available, without waiting for subsequent QEC rounds outside that window. After an A window has been decoded, recovery contributions supported entirely within its core are retained, whereas seam-crossing matching information is encoded as artificial boundary defects and passed as boundary information to the corresponding B window. The B-window task $\text{B}_i$ becomes eligible for execution once its two adjacent A-window tasks, $\text{A}_i$ and $\text{A}_{i+1}$, have completed and provided the required boundary information. This dependency structure permits a ready B-window task to execute without waiting for unrelated A-window tasks, thereby enabling decoding to overlap syndrome acquisition and transmission.
Once the final round of syndromes has been received by the host, all remaining window tasks whose input dependencies are satisfied can be executed. The window-decoding stage is complete when all A- and B-window tasks required for the current decoding instance have finished.
Finally, the recovery contributions produced by the completed A and B windows are combined so that duplicate contributions cancel. 
The resulting correction (i.e., corrective Pauli operators) are sent back along the reverse path to the FPGA correction logic, which will trigger the quantum control system for applying corresponding corrections on QPUs. The Switch decouples FPGA syndrome I/O from host decoder placement, enabling the system to scale out by attaching additional FPGA and HPC nodes through TH-Express ports.

We measure the end-to-end decoding latency as
$\tau_{\mathrm{e2e}}
    =
    t_{\mathrm{corr}}
    -
    t_{\mathrm{syn,last}}$,
where $t_{\mathrm{syn,last}}$ denotes the time at which transmission of the final round of syndromes starts, and $t_{\mathrm{corr}}$ denotes the time at which the complete correction set $\mathcal{C}$ has been received and made available at the FPGA correction module. Both timestamps are recorded in the same FPGA clock domain. 
This interval includes transmission of the syndromes to the host, the decoding time, correction calculation and mapping, and transmission of the correction back to the FPGA. 
Syndrome processing completed before $t_{\mathrm{syn,last}}$ is hidden by its overlap with syndrome acquisition and streaming. 

Software-level decoding latency is evaluated separately using a batch method. For each sample, the complete syndrome streams are available in host memory before timing begins. The parallel-window decoder and the global decoder (that decodes all syndromes at once) are evaluated using identical syndrome instances and same timing strategies. Syndrome generation and offline initialization are excluded for software-level decoding latency evaluation.

\section{Results}
\label{sec:results}

\subsection{Experiment setup}
\label{sec:setup}
We conduct experiments on Intel Xeon Gold 6348 CPUs running at 2.60GHz, featuring 28 physical cores and 56 logical processors, with 43,008 KB of cache, and a 46 bits physical and 57 bits virtual address space architecture.The server is equipped with 512GB of memory. We also utilize an AMD Virtex UltraScale+ FPGA in our experimental setup. The FPGA emulates the quantum control system, while the HPC cluster performs syndrome decoding and returns corrections through TH-Express. This high-performance computing environment provides the necessary computational resources for our experimental evaluations.
We use the MWPM decoder, Sparse blossom~\cite{higgott2025sparse} as our base decoder. We evaluate the decoding latency under the circuit-level
depolarizing noise model, where each single-qubit or two-qubit gate has depolarising noise with probability $p$, each reset or measurement flips with probability $p$.
We first characterize the communication latency of TH-Express, which constitutes the communication component of the end-to-end decoding latency. We then evaluate the complete end-to-end decoding pipeline on the surface-code memory experiments using both global and parallel-window decoding strategies.

\subsection{Network latency characterization}
\label{sec:res-network}

\begin{figure}[tbp]
\centering
\includegraphics[width=\linewidth]{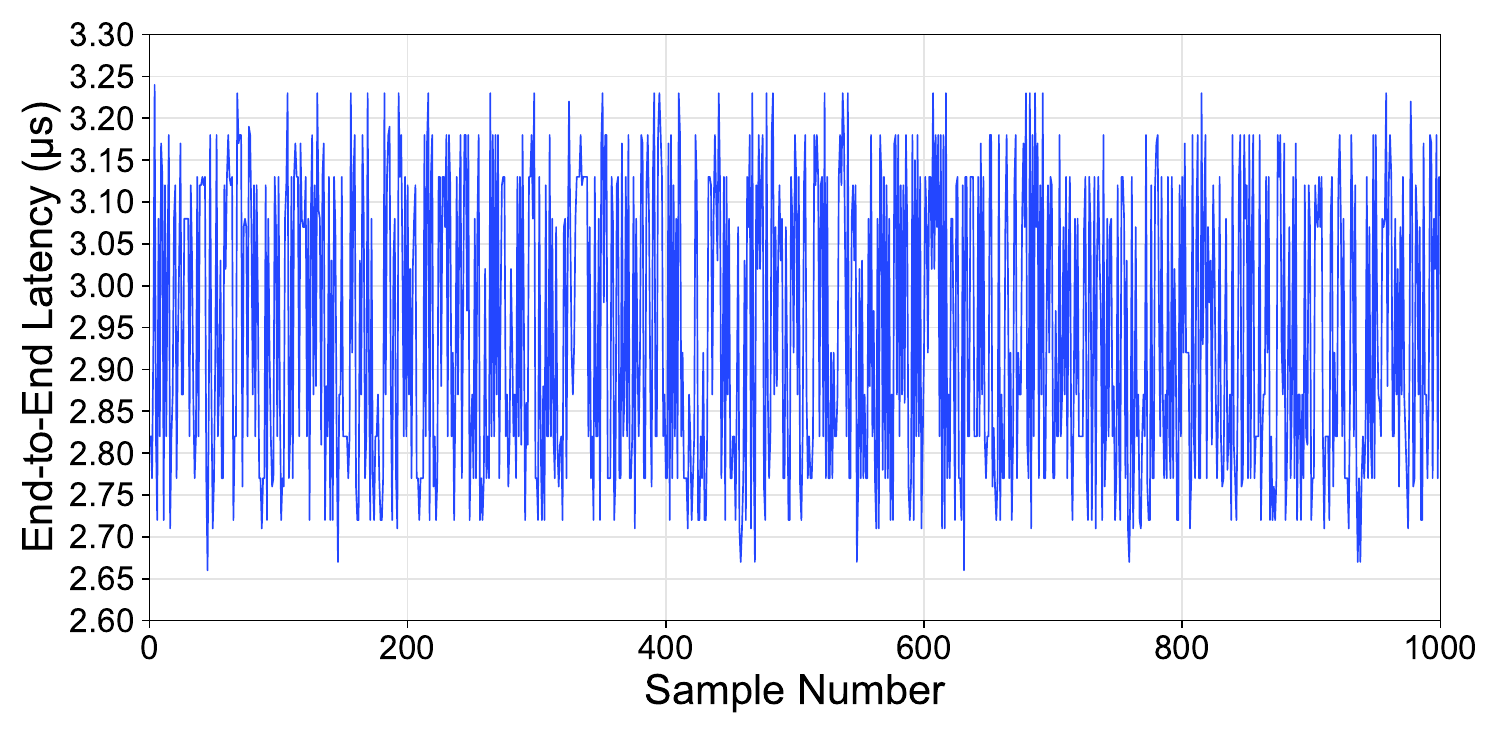}
\includegraphics[width=\linewidth]{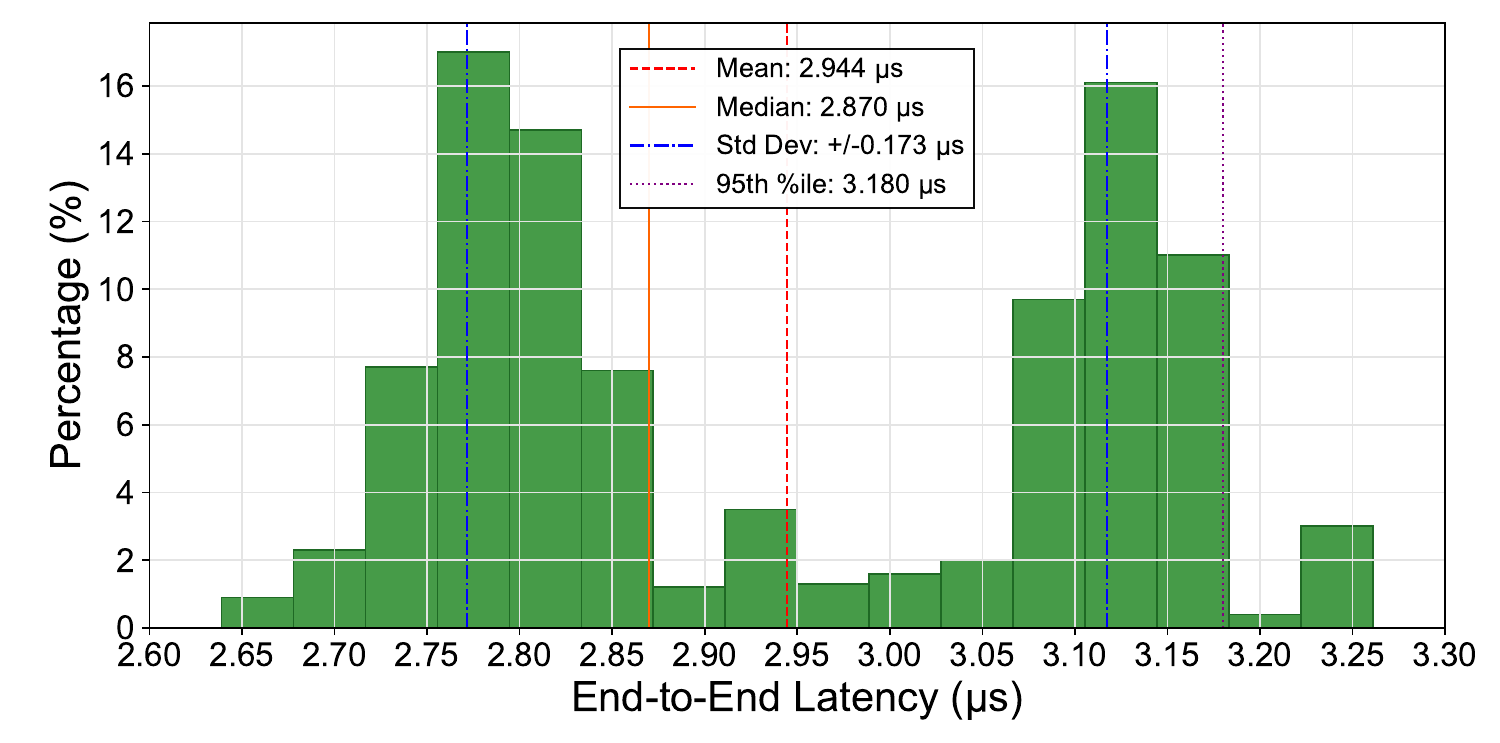}
\caption{The round latency of TH-Express with direct interconnect between HPC and FPGA.}
\label{fig:th-latency}
\end{figure}

\begin{figure}[tbp!]
    \centering
    \includegraphics[width=1\linewidth]{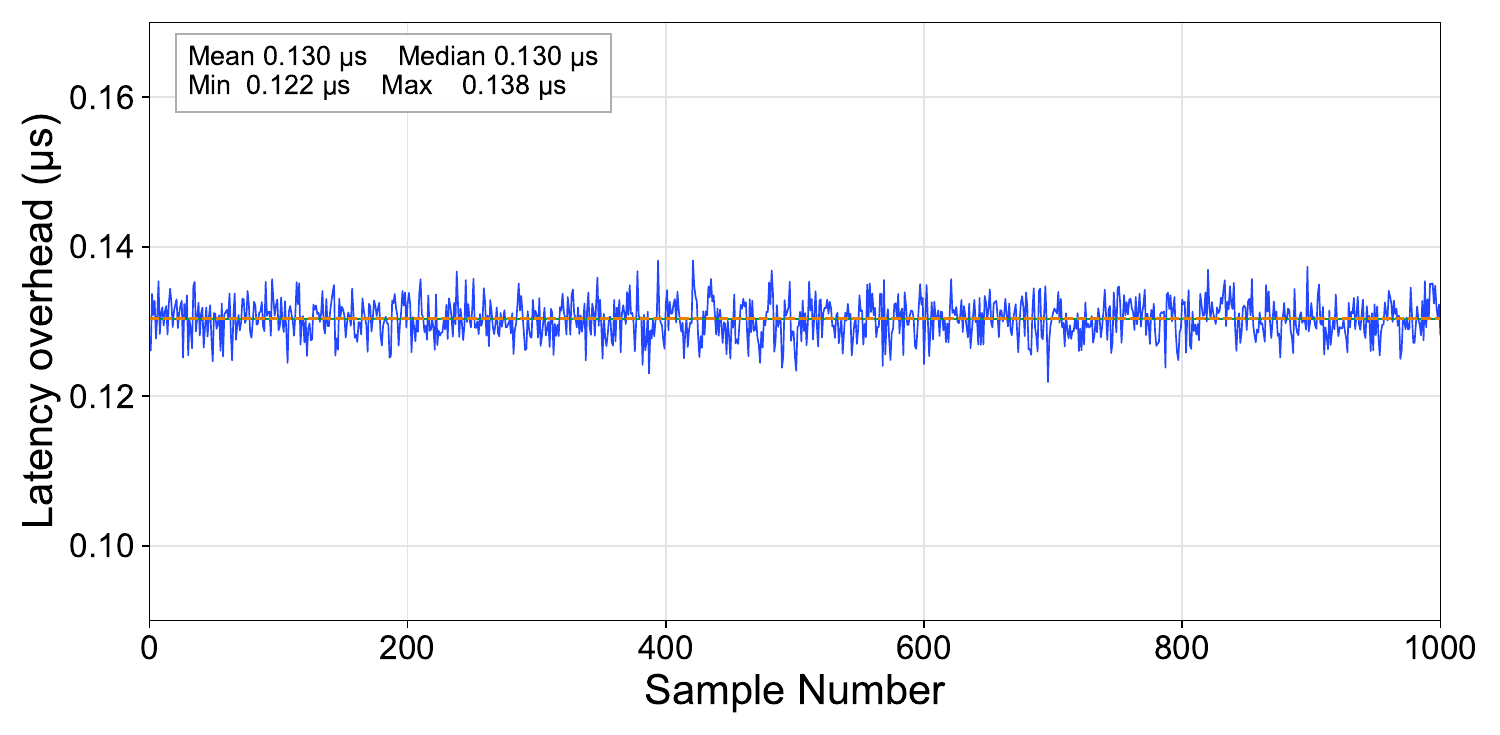}
    \caption{Latency overhead per additional hop}
    \label{fig:thlink-switch}
\end{figure}

Since communication latency directly contributes to the end-to-end decoding response time, we first quantify the raw performance of the TH-Express interconnect under decoding traffic. We measure the round-trip latency of direct interconnect between the FPGA and one HPC rack in Figure~\ref{fig:th-latency}.
The mean and median round-trip latencies are 2.944~$\mu$s and 2.870~$\mu$s respectively, with a standard deviation of 173~ns, a minimum of 2.66~$\mu$s, and a maximum of 3.24~$\mu$s.
Figure~\ref{fig:thlink-switch} shows the incremental latency introduced by each additional switch hop when scaling to more computing nodes, the mean and median latencies are 0.130~$\mu$s and 0.130~$\mu$s respectively, with a minimum of 0.122~$\mu$s, and a maximum of 0.138~$\mu$s.

\begin{figure}[tbp!]
    \centering
    \includegraphics[width=0.8\linewidth]{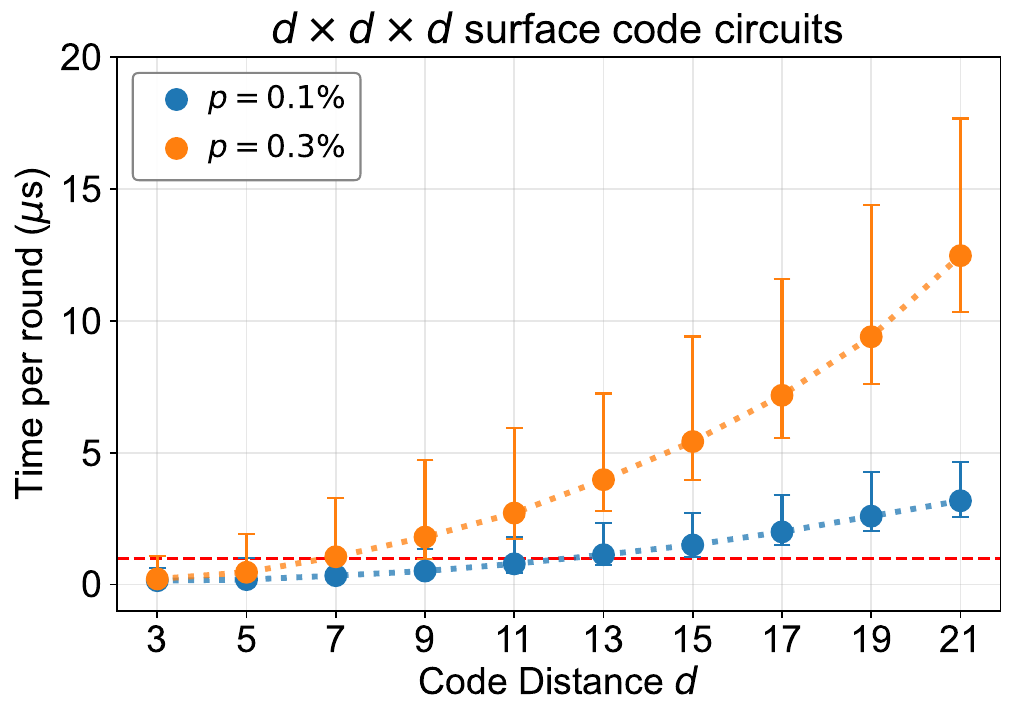}
    \caption{End-to-end decoding latency for $d \times d \times d$ surface-code memory circuits ($p=0.1\%$) using global MWPM. Each point shows the median over independent syndrome instances.}
    \label{fig:ee-dl-d-mwpm}
\end{figure}

\begin{figure*}[htb!]
    \centering
    \includegraphics[width=0.35\linewidth]{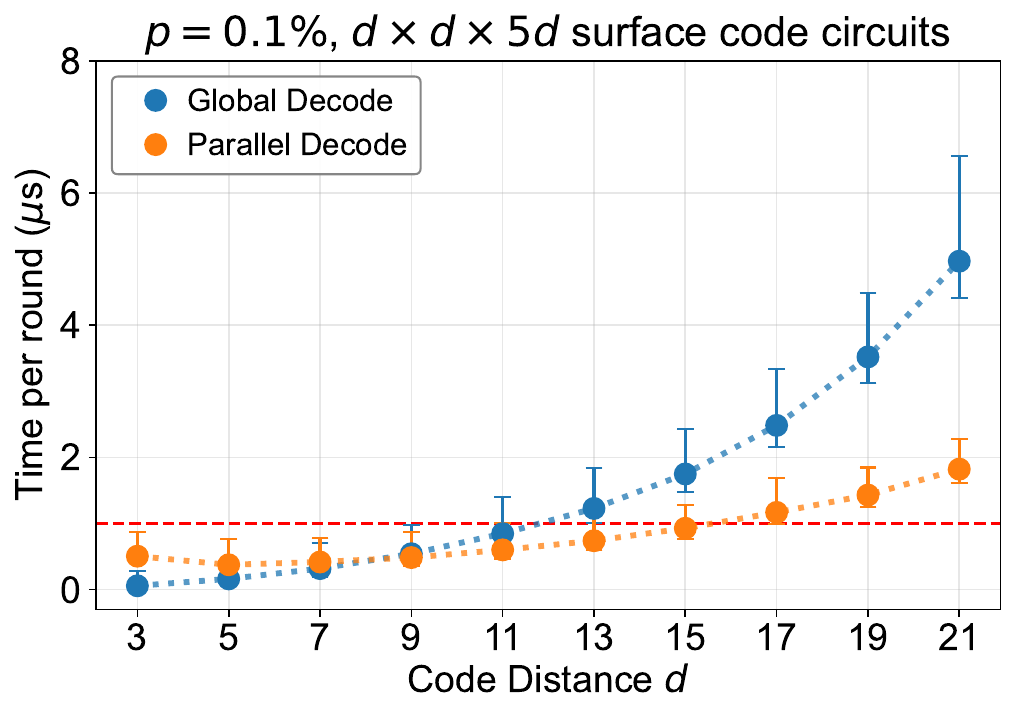}
    \includegraphics[width=0.35\linewidth]{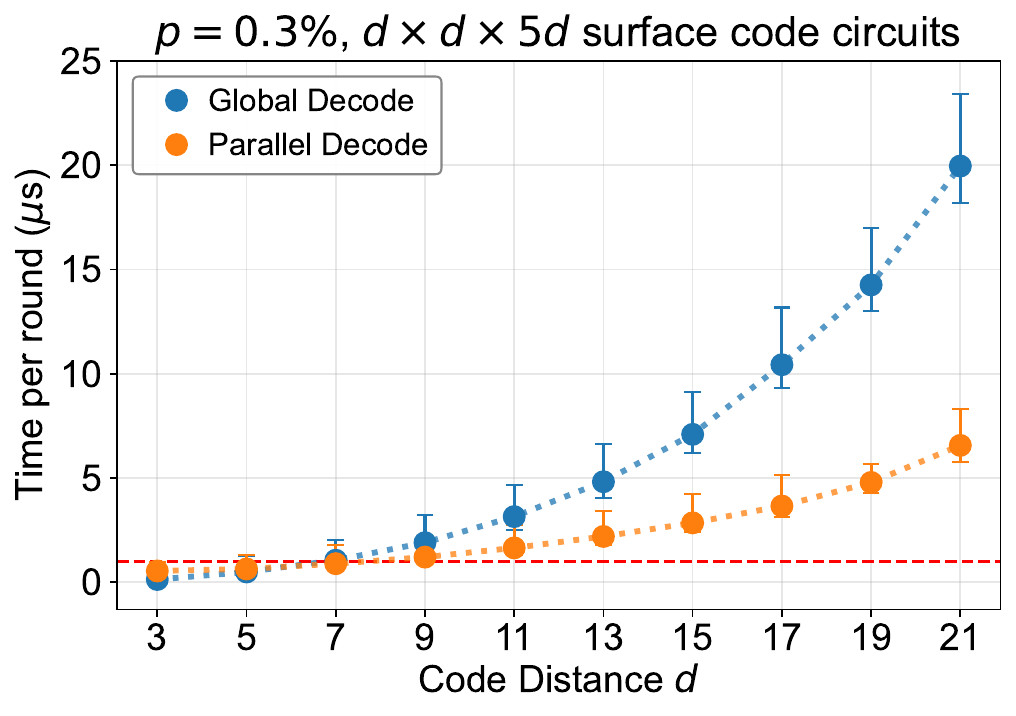}
    \includegraphics[width=0.35\linewidth]{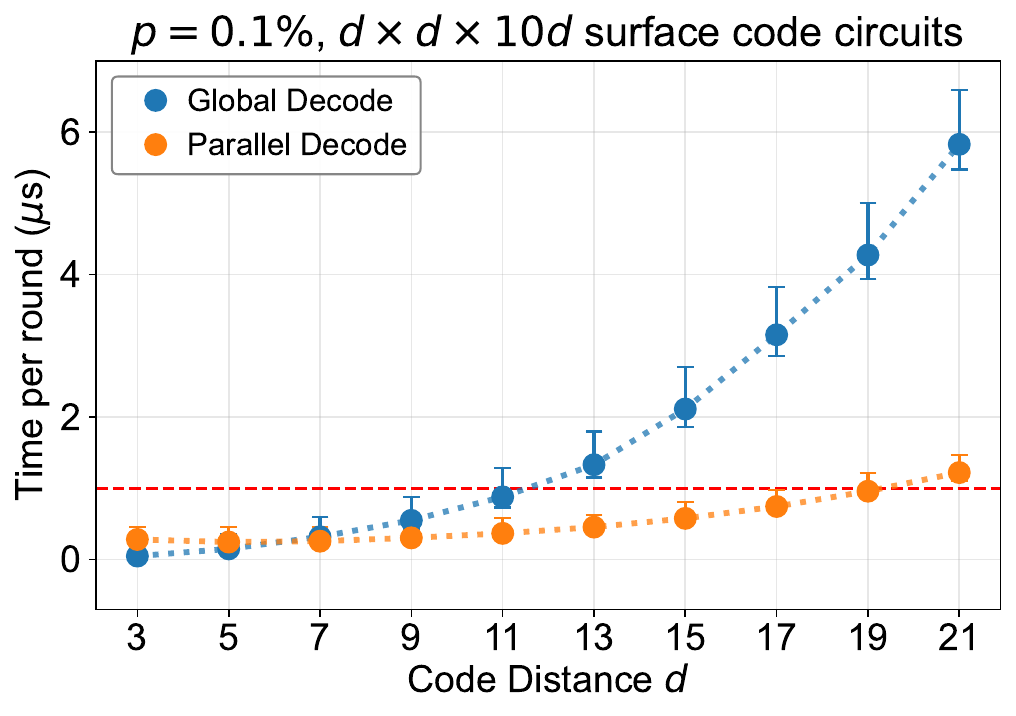}
    \includegraphics[width=0.35\linewidth]{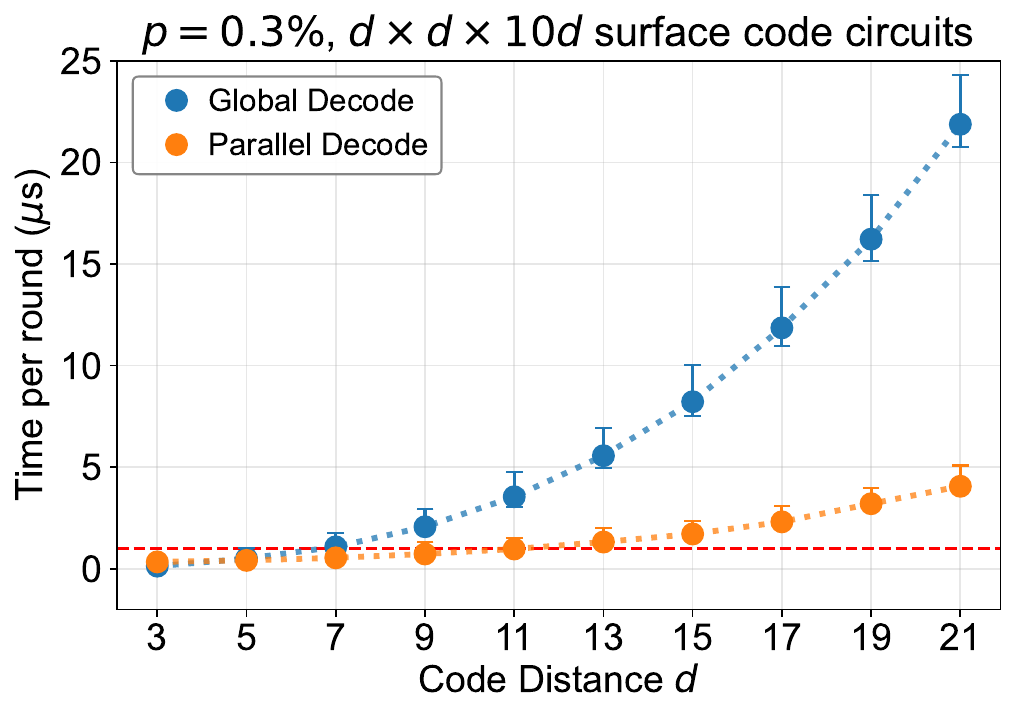}
    \caption{End-to-end decoding latency for $d \times d \times 5d$ (top) and $d \times d \times 10d$ (bottom) surface-code memory circuits using global MWPM and parallel window with buffer size $b=(d+1)/2$.}
    \label{fig:ee-dl-nd-mwpm}
\end{figure*}
\subsection{End-to-end decoding latency}
\label{sec:res-decoding}

We evaluate the end-to-end decoding latency for surface-code memory experiments running over the \emph{THQLink} architecture. 
During the execution of quantum error correction, the FPGA streams round-resolved syndrome data through TH-Express to the HPC system round by round as the syndrome measurements are generated. 

Figure~\ref{fig:ee-dl-d-mwpm} reports the end-to-end latency for decoding $d$ rounds of syndromes (i.e., $d \times d \times d$) at physical error rates $p=0.1\%$ and $p=0.3\%$ using global MWPM. 
For physical error rate $p=0.1\%$, the \emph{THQLink} architecture could implement real-time decoding for the distance-13 surface code.
Interconnect latency limits the largest code distance that can be supported in real-time decoding.
Figure~\ref{fig:ee-dl-nd-mwpm} compares the end-to-end latency of global MWPM against parallel window decoding for $5d$ rounds (i.e., $d \times d \times 5d$) and $1d$ rounds (i.e., $d \times d \times 1d$)  memories, using window step size $s=(d+1)/2$ and buffer size $b=(d+1)/2$. 
The syndrome stream is partitioned across multiple cores in the HPC rack, enabling concurrent window decoding. 
At small distances ($d \le 7$), parallel window carries overhead from window construction, seam stitching, and the additional network traffic for boundary information, so global MWPM retains a marginal advantage. 
As $d$ increases, the parallel-window decoding amortizes its overhead and outperforms the global baseline decoding. 
For $\times d \times 5d$ at $p=0.1\%$, the parallel window achieves real-time decoding for distance 15, while the global decoder reaches the time budget at distance 11.
Real-time decoding for larger distances can be achieved by increasing the syndrome rounds. For $d \times d \times 10d$ (bottom)  at $p=0.1\%$, the parallel window achieves real-time decoding for distance 19, while the global decoder reaches the time budget at distance 11.
At the higher physical error rate ($p=0.3\%$), denser syndrome graphs increase the absolute latency for both strategies because more matching iterations are required.

\section{Discussion and conclusion}
\label{sec:conclusion}

This work introduced THQLink, a scalable real‑time decoding architecture that tightly couples a general‑purpose HPC cluster to a quantum control system via a low‑latency TH‑Express network~\cite{xu2020scalable}. Our measurements show a mean round‑trip latency of 2.944~$\mu$s and an incremental cost of only 130~ns per additional hop. Using a parallel window decoding strategy, we demonstrated real‑time decoding of surface codes for up to distance 19 using a matching-based decoder on CPUs. These results demonstrate that a programmable HPC system with a high‑speed interconnect can deliver real‑time performance while preserving algorithmic flexibility and scalability.

THQLink allows seamless adoption of emerging QEC codes and decoders without costly redesign cycles. The measured incremental latency suggests that this architecture can scale to a large number of logical qubits provided the network topology and workload distribution are carefully optimized. THQLink is also suited for quantum-centric supercomputers, where tight coupling between QPU and HPC resources enables efficient support for hybrid quantum-classical applications.
Future work includes validating the architecture with qLDPC codes, developing adaptive load‑balancing mechanisms for heterogeneous HPC nodes.

\begin{acknowledgments}
We acknowledge the support from QUANTA (QUantum fANs from iT Area) group. This work has been supported by the National Key R\&D Program of China (Grant No. 2024YFB4504001), the National Natural Science Foundation of China (Grant No. 62302395 and 62421002), the Fundamental and Interdisciplinary Disciplines Breakthrough Plan of the Ministry of Education of China (Grant No. JYB2025XDXM202), the Aid Program for Science and Technology Innovative Research Team in Higher Educational Institutions of Hunan Province. 
\end{acknowledgments}

\bibliography{references}

@PREAMBLE{
 "\providecommand{\noopsort}[1]{}" 
 # "\providecommand{\singleletter}[1]{#1}%" 
}

@article{shor1995scheme,
  title={Scheme for reducing decoherence in quantum computer memory},
  author={Shor, Peter W},
  journal={Physical review A},
  volume={52},
  number={4},
  pages={R2493},
  year={1995},
  publisher={APS},
}

@inproceedings{shor1996fault,
  title={Fault-tolerant quantum computation},
  author={Shor, Peter W},
  booktitle={Proceedings of 37th conference on foundations of computer science},
  pages={56--65},
  year={1996},
  organization={IEEE},
}

@article{calderbank1996good,
  title={Good quantum error-correcting codes exist},
  author={Calderbank, A Robert and Shor, Peter W},
  journal={Physical Review A},
  volume={54},
  number={2},
  pages={1098},
  year={1996},
  publisher={APS},
}

@article{terhal2015quantum,
  title={Quantum error correction for quantum memories},
  author={Terhal, Barbara M},
  journal={Reviews of Modern Physics},
  volume={87},
  number={2},
  pages={307},
  year={2015},
  publisher={APS},
}

@article{kitaev2003fault,
  title={Fault-tolerant quantum computation by anyons},
  author={Kitaev, A Yu},
  journal={Annals of physics},
  volume={303},
  number={1},
  pages={2--30},
  year={2003},
  publisher={Elsevier},
}

@article{bravyi1998quantum,
  title={Quantum codes on a lattice with boundary},
  author={Bravyi, Sergey B and Kitaev, A Yu},
  journal={quant-ph/9811052},
  year={1998}
}

@article{dennis2002topological,
  title={Topological quantum memory},
  author={Dennis, Eric and Kitaev, Alexei and Landahl, Andrew and Preskill, John},
  journal={Journal of Mathematical Physics},
  volume={43},
  number={9},
  pages={4452--4505},
  year={2002},
  publisher={American Institute of Physics},
}

@article{fowler2012surface,
  title = {Surface codes: Towards practical large-scale quantum computation},
  author = {Fowler, Austin G. and Mariantoni, Matteo and Martinis, John M. and Cleland, Andrew N.},
  journal = {Phys. Rev. A},
  volume = {86},
  issue = {3},
  pages = {032324},
  numpages = {48},
  year = {2012},
  month = {Sep},
  publisher = {American Physical Society}
}

@article{bombin2006topological,
  title = {Topological Quantum Distillation},
  author = {Bombin, H. and Martin-Delgado, M. A.},
  journal = {Phys. Rev. Lett.},
  volume = {97},
  issue = {18},
  pages = {180501},
  numpages = {4},
  year = {2006},
  month = oct,
  publisher = {American Physical Society},
}

@article{bombin2015gauge,
  title={Gauge color codes: optimal transversal gates and gauge fixing in topological stabilizer codes},
  author={Bomb{\'\i}n, H{\'e}ctor},
  journal={New Journal of Physics},
  volume={17},
  number={8},
  pages={083002},
  year={2015},
  publisher={IOP Publishing}
}

@article{kubica2015universal,
  title={Universal transversal gates with color codes: A simplified approach},
  author={Kubica, Aleksander and Beverland, Michael E},
  journal={Physical Review A},
  volume={91},
  number={3},
  pages={032330},
  year={2015},
  publisher={APS}
}

@book{kubica2018abcs,
  title={The ABCs of the color code: A study of topological quantum codes as toy models for fault-tolerant quantum computation and quantum phases of matter},
  author={Kubica, Aleksander Marek},
  year={2018},
  publisher={California Institute of Technology}
}

@article{higgott2025sparse,
  title={Sparse blossom: correcting a million errors per core second with minimum-weight matching},
  author={Higgott, Oscar and Gidney, Craig},
  journal={Quantum},
  volume={9},
  pages={1600},
  year={2025},
  publisher={Verein zur F{\"o}rderung des Open Access Publizierens in den Quantenwissenschaften}
}

@inproceedings{wu2023fusion,
  title={Fusion blossom: Fast mwpm decoders for qec},
  author={Wu, Yue and Zhong, Lin},
  booktitle={2023 IEEE International Conference on Quantum Computing and Engineering (QCE)},
  volume={1},
  pages={928--938},
  year={2023},
  organization={IEEE}
}

@article{delfosse2021almost,
  title={Almost-linear time decoding algorithm for topological codes},
  author={Delfosse, Nicolas and Nickerson, Naomi H},
  journal={Quantum},
  volume={5},
  pages={595},
  year={2021},
  publisher={Verein zur F{\"o}rderung des Open Access Publizierens in den Quantenwissenschaften}
}

@article{higgott2022pymatching,
  title={Pymatching: A python package for decoding quantum codes with minimum-weight perfect matching},
  author={Higgott, Oscar},
  journal={ACM Transactions on Quantum Computing},
  volume={3},
  number={3},
  pages={1--16},
  year={2022},
  publisher={ACM New York, NY}
}

@article{tan2023,
  title={Scalable Surface-Code Decoders with Parallelization in Time},
  author={Tan, Xinyu and Zhang, Fang and Chao, Rui and Shi, Yaoyun and Chen, Jianxin},
  journal={PRX Quantum},
  volume={4},
  pages={040344},
  year={2023},
  publisher={APS}
}

@article{dennis2002,
  title={Topological quantum memory},
  author={Dennis, Eric and Kitaev, Alexei and Landahl, Andrew and Preskill, John},
  journal={Journal of Mathematical Physics},
  volume={43},
  pages={4452--4505},
  year={2002},
  publisher={American Institute of Physics}
}

@article{das2022,
  title={A scalable decoder for error correction of quantum computing},
  author={Das, Poulami and Locharla, Animesh and Jones, Cody},
  journal={Proceedings of the 27th ACM International Conference on Architectural Support for Programming Languages and Operating Systems},
  pages={541--555},
  year={2022},
  publisher={ACM}
}

@article{skoric2023,
  title={Parallel window decoding enables scalable fault tolerant quantum computation},
  author={Skoric, Luka and Browne, Daniel E and Barnes, Kieran M and Gillespie, Nathan I and Campbell, Earl T},
  journal={Nature Communications},
  volume={14},
  pages={7040},
  year={2023},
  publisher={Nature Publishing Group}
}

@article{Poulin2008iterative,
  title={On the iterative decoding of sparse quantum codes},
  author={Poulin, David and Chung, Yeojin},
  journal={arXiv preprint arXiv:0801.1241},
  year={2008}
}

@article{Muller2025-tt,
  title={Improved belief propagation is sufficient for real-time
                   decoding of quantum memory},
  author={M{\"u}ller, Tristan and Alexander, Thomas and Beverland,
                   Michael E and B{\"u}hler, Markus and Johnson, Blake R and
                   Maurer, Thilo and Vandeth, Drew},
  journal={arXiv preprint arXiv:2506.01779},
  year={2025}
}

@article{barber2025real,
  title={A real-time, scalable, fast and resource-efficient decoder for a quantum computer},
  author={Barber, Ben and Barnes, Kenton M and Bialas, Tomasz and Bu{\u{g}}dayc{\i}, Okan and Campbell, Earl T and Gillespie, Neil I and Johar, Kauser and Rajan, Ram and Richardson, Adam W and Skoric, Luka and others},
  journal={Nature Electronics},
  volume={8},
  number={1},
  pages={84--91},
  year={2025},
  publisher={Nature Publishing Group UK London}
}

@article{caldwell2025platform,
  title={Platform Architecture for Tight Coupling of High-Performance Computing with Quantum Processors},
  author={Caldwell, Shane A and Khazraee, Moein and Agostini, Elena and Lassiter, Tom and Simpson, Corey and Kahalon, Omri and Kanuri, Mrudula and Kim, Jin-Sung and Stanwyck, Sam and Li, Muyuan and others},
  journal={arXiv preprint arXiv:2510.25213},
  year={2025}
}

@article{maurer2025real,
  title={Real-time decoding of the gross code memory with FPGAs},
  author={Maurer, Thilo and B{\"u}hler, Markus and Kr{\"o}ner, Michael and Haverkamp, Frank and M{\"u}ller, Tristan and Vandeth, Drew and Johnson, Blake R},
  journal={arXiv preprint arXiv:2510.21600},
  year={2025}
}

@article{maurya2025fpga,
  title={FPGA-tailored algorithms for real-time decoding of quantum LDPC codes},
  author={Maurya, Satvik and Maurer, Thilo and B{\"u}hler, Markus and Vandeth, Drew and Beverland, Michael E},
  journal={arXiv preprint arXiv:2511.21660},
  year={2025}
}

@article{olle2024simultaneous,
  title={Simultaneous discovery of quantum error correction codes and encoders with a noise-aware reinforcement learning agent},
  author={Olle, Jan and Zen, Remmy and Puviani, Mattia and Marquardt, Florian},
  journal={npj Quantum Information},
  volume={10},
  pages={136},
  year={2024},
  publisher={Nature Publishing Group}
}

@article{sweke2021reinforcement,
  title={Reinforcement learning decoders for fault-tolerant quantum computation},
  author={Sweke, Ryan and Kesselring, Markus S and van Nieuwenburg, Evert PL and Eisert, Jens},
  journal={Machine Learning: Science and Technology},
  volume={2},
  number={2},
  pages={025005},
  year={2021},
  publisher={IOP Publishing}
}

@article{matekole2022decoding,
  title={Decoding surface codes with deep reinforcement learning and probabilistic policy reuse},
  author={Matekole, Ephraim S and Ye, Emily and Iyer, Rahul and Chen, Samuel Yen-Chi},
  journal={arXiv preprint arXiv:2212.11890},
  year={2022}
}

@article{he2025discovering,
  title={Discovering highly efficient low-weight quantum error-correcting codes with reinforcement learning},
  author={He, Andrew Y and Liu, Zi-Wen},
  journal={arXiv preprint arXiv:2502.14372},
  year={2025}
}

@article{lange2025datadriven,
  title={Data-driven decoding of quantum error correcting codes using graph neural networks},
  author={Lange, Moritz and Havstr{\"o}m, Pontus and Srivastava, Basudha and Bengtsson, Isak and Bergentall, Valdemar and Hammar, Karl and Heuts, Olivia and van Nieuwenburg, Evert and Granath, Mats},
  journal={Physical Review Research},
  volume={7},
  number={2},
  pages={023181},
  year={2025},
  publisher={American Physical Society}
}

@inproceedings{gong2024graph,
  title={Graph neural networks for enhanced decoding of quantum {LDPC} codes},
  author={Gong, Anqi and Cammerer, Sebastian and Renes, Joseph M},
  booktitle={2024 IEEE International Symposium on Information Theory (ISIT)},
  pages={2700--2705},
  year={2024},
  organization={IEEE}
}

@article{maan2025machine,
  title={Machine learning message-passing for the scalable decoding of {QLDPC} codes},
  author={Maan, Arshpreet Singh and Paler, Alexandru},
  journal={npj Quantum Information},
  volume={11},
  pages={78},
  year={2025},
  publisher={Nature Publishing Group}
}

@article{liu2025decoding,
  title={Decoding quantum low density parity check codes with diffusion},
  author={Liu, Zejun and Gong, Anqi and Clark, Bryan K},
  journal={arXiv preprint arXiv:2509.22347},
  year={2025}
}

@article{cao2025generative,
  title={Generative decoding for quantum error-correcting codes},
  author={Cao, Hao and Pan, Feng and Feng, Dong and Wang, Yuzhen and Zhang, Pan},
  journal={arXiv preprint arXiv:2503.21374},
  year={2025}
}

@article{bausch2024learning,
  title={Learning high-accuracy error decoding for quantum processors},
  author={Bausch, Johannes and Senior, Andrew W and Heras, Francisco JH and Edlich, Thomas and Davies, Alex and Newman, Michael and Jones, Cody and Satzinger, Kevin and Niu, Murphy Yuezhen and Blackwell, Sam and others},
  journal={Nature},
  volume={635},
  number={8040},
  pages={834--840},
  year={2024},
  publisher={Nature Publishing Group}
}

@inproceedings{choukroun2022error,
  title={Error correction code transformer},
  author={Choukroun, Yoni and Wolf, Lior},
  booktitle={Advances in Neural Information Processing Systems (NeurIPS)},
  volume={35},
  pages={23390--23402},
  year={2022}
}

@article{cohen2025hybrid,
  title={Hybrid mamba-transformer decoder for error-correcting codes},
  author={Cohen, Shy-el and Choukroun, Yoni and Nachmani, Eliya},
  journal={arXiv preprint arXiv:2505.17834},
  year={2025}
}

@article{steane1996error,
  title={Error correcting codes in quantum theory},
  author={Steane, Andrew M},
  journal={Physical Review Letters},
  volume={77},
  number={5},
  pages={793},
  year={1996},
  publisher={APS}
}

@book{gottesman1997stabilizer,
  title={Stabilizer codes and quantum error correction},
  author={Gottesman, Daniel},
  year={1997},
  publisher={California Institute of Technology}
}

@article{google2025quantum,
  title={Quantum error correction below the surface code threshold},
  author={Google Quantum AI and Collaborators},
  journal={Nature},
  volume={638},
  number={8052},
  pages={920--926},
  year={2025},
  publisher={Nature Publishing Group UK London}
}

@article{skoric2023parallel,
  title={Parallel window decoding enables scalable fault tolerant quantum computation},
  author={Skoric, Luka and Browne, Dan E and Barnes, Kenton M and Gillespie, Neil I and Campbell, Earl T},
  journal={Nature Communications},
  volume={14},
  number={1},
  pages={7040},
  year={2023},
  publisher={Nature Publishing Group UK London}
}

@article{caune2026demonstrating,
  title={Demonstrating real-time and low-latency quantum error correction with superconducting qubits},
  author={Caune, Laura and Skoric, Luka and Blunt, Nick S and Ruban, Archibald and McDaniel, Jimmy and Valery, Joseph A and Patterson, Andrew D and Gramolin, Alexander V and Majaniemi, Joonas and Barnes, Kenton M and others},
  journal={Nature Communications},
  year={2026},
  publisher={Nature Publishing Group UK London}
}

@book{nielsen2010quantum,
  title={Quantum computation and quantum information},
  author={Nielsen, Michael A and Chuang, Isaac L},
  year={2010},
  publisher={Cambridge university press}
}

@article{bravyi2024high,
  title={High-threshold and low-overhead fault-tolerant quantum memory},
  author={Bravyi, Sergey and Cross, Andrew W and Gambetta, Jay M and Maslov, Dmitri and Rall, Patrick and Yoder, Theodore J},
  journal={Nature},
  volume={627},
  number={8005},
  pages={778--782},
  year={2024},
  publisher={Nature Publishing Group UK London}
}

@article{bluvstein2026fault,
  title={A fault-tolerant neutral-atom architecture for universal quantum computation},
  author={Bluvstein, Dolev and Geim, Alexandra A and Li, Sophie H and Evered, Simon J and Bonilla Ataides, J Pablo and Baranes, Gefen and Gu, Andi and Manovitz, Tom and Xu, Muqing and Kalinowski, Marcin and others},
  journal={Nature},
  volume={649},
  number={8095},
  pages={39--46},
  year={2026},
  publisher={Nature Publishing Group UK London}
}

@article{zhao2026towards,
  title={Towards ultra-high-rate quantum error correction with reconfigurable atom arrays},
  author={Zhao, Chen and Duckering, Casey and Gu, Andi and Maskara, Nishad and Zhou, Hengyun},
  journal={arXiv preprint arXiv:2604.16209},
  year={2026}
}

@article{ziad2025local,
  title={Local clustering decoder as a fast and adaptive hardware decoder for the surface code},
  author={Ziad, Abbas B and Zalawadiya, Ankit and Topal, Canberk and Camps, Joan and Geh{\'e}r, Gy{\"o}rgy P and Stafford, Matthew P and Turner, Mark L},
  journal={Nature Communications},
  volume={16},
  number={1},
  pages={11048},
  year={2025},
  publisher={Nature Publishing Group UK London}
}

@article{senior2025scalable,
  title={A scalable and real-time neural decoder for topological quantum codes},
  author={Senior, Andrew W and Edlich, Thomas and Heras, Francisco JH and Zhang, Lei M and Higgott, Oscar and Spencer, James S and Applebaum, Taylor and Blackwell, Sam and Ledford, Justin and {\v{Z}}emgulyt{\.e}, Akvil{\.e} and others},
  journal={arXiv preprint arXiv:2512.07737},
  year={2025}
}

@article{tan2023scalable,
  title={Scalable surface-code decoders with parallelization in time},
  author={Tan, Xinyu and Zhang, Fang and Chao, Rui and Shi, Yaoyun and Chen, Jianxin},
  journal={PRX Quantum},
  volume={4},
  number={4},
  pages={040344},
  year={2023},
  publisher={APS}
}

@article{paetznick2026improved,
  title={Improved quantum processor logical error rates via correction and detection},
  author={Paetznick, A and Reichardt, BW and da Silva, MP and Ryan-Anderson, C and Aasen, D and Bello-Rivas, JM and Campora, JP and Chao, R and Chernoguzov, A and van Dam, W and others},
  journal={Nature},
  volume={654},
  number={8118},
  pages={349--355},
  year={2026},
  publisher={Nature Publishing Group}
}

@article{xu2020scalable,
  title={A scalable smart router architecture with intelligent adaptive routing and fault-tolerant management},
  author={Xu, Shi and Lai, Mingche and Dai, Yi and Cao, Jijun and Wang, Kefei},
  journal={Neurocomputing},
  volume={393},
  pages={126--141},
  year={2020},
  publisher={Elsevier}
}

@article{xie2011tianhe,
  title={Tianhe-1a interconnect and message-passing services},
  author={Xie, Min and Lu, Yutong and Wang, Kefei and Liu, Lu and Cao, Hongjia and others},
  journal={IEEE micro},
  volume={32},
  number={1},
  pages={8--20},
  year={2011},
  publisher={IEEE}
}

@article{zhang2026learning,
  title={Learning to Decode in Parallel: Self-Coordinating Neural Network for Real-Time Quantum Error Correction},
  author={Zhang, Kai and Yi, Zhengzhong and Guo, Shaojun and Kong, Linghang and Wang, Situ and Zhan, Xiaoyu and He, Tan and Lin, Weiping and Jiang, Tao and Gao, Dongxin and others},
  journal={arXiv preprint arXiv:2601.09921},
  year={2026}
}

@inproceedings{wu2025micro,
  title={Micro blossom: Accelerated minimum-weight perfect matching decoding for quantum error correction},
  author={Wu, Yue and Liyanage, Namitha and Zhong, Lin},
  booktitle={Proceedings of the 30th ACM International Conference on Architectural Support for Programming Languages and Operating Systems, Volume 2},
  pages={639--654},
  year={2025}
}

@article{das2020scalable,
  title={A scalable decoder micro-architecture for fault-tolerant quantum computing},
  author={Das, Poulami and Pattison, Christopher A and Manne, Srilatha and Carmean, Douglas and Svore, Krysta and Qureshi, Moinuddin and Delfosse, Nicolas},
  journal={arXiv preprint arXiv:2001.06598},
  year={2020}
}
\newpage
\appendix

\section{Parallel window decoding}
\label{app:parallel}

\begin{figure*}[htb!]
    \centering
    \includegraphics[width=0.6\linewidth]{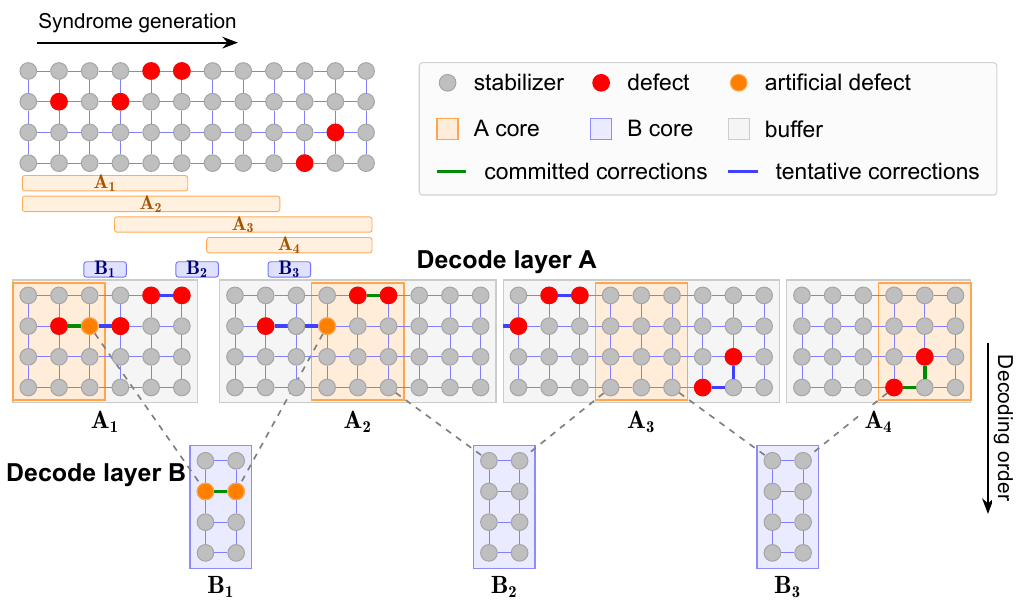}
    \caption{The parallel window scheme}
    \label{fig:parallel-window}
\end{figure*}

\begin{figure}[tbh!]
    \centering
    \includegraphics[width=0.8\linewidth]{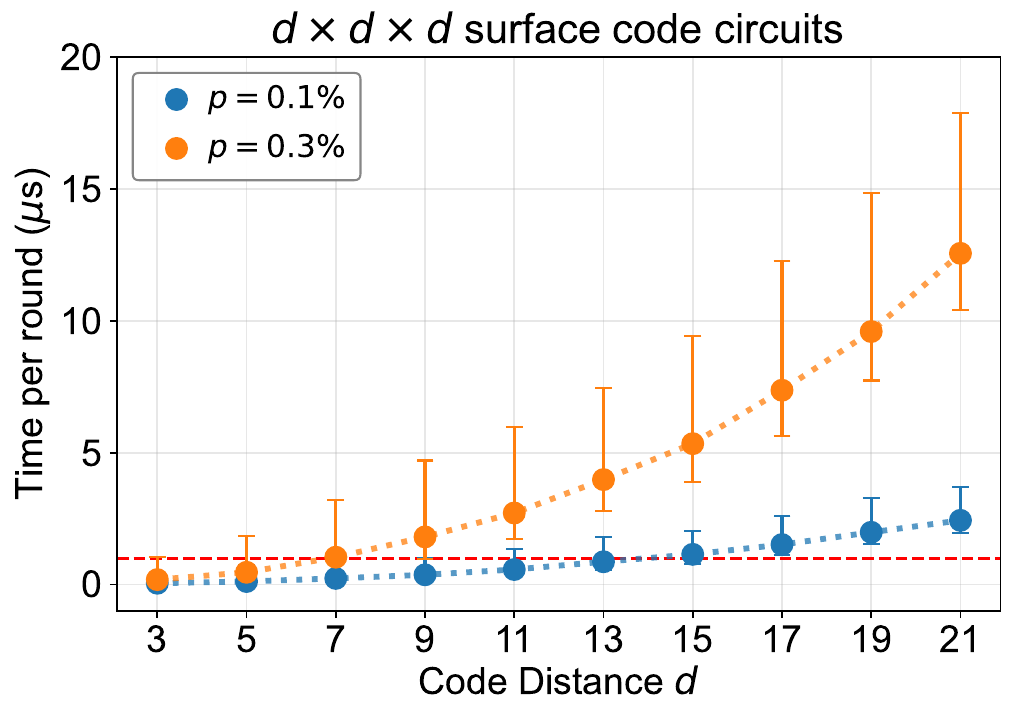}
    \caption{Software-level decoding latency per round for $d \times d \times d$ surface-code memory circuits ($p=0.1\%$) using global MWPM. Each point shows the median over independent syndrome instances.}
    \label{fig:soft-dl-d-mwpm}
\end{figure}

\begin{figure*}[tbh!]
    \centering
    \includegraphics[width=0.35\linewidth]{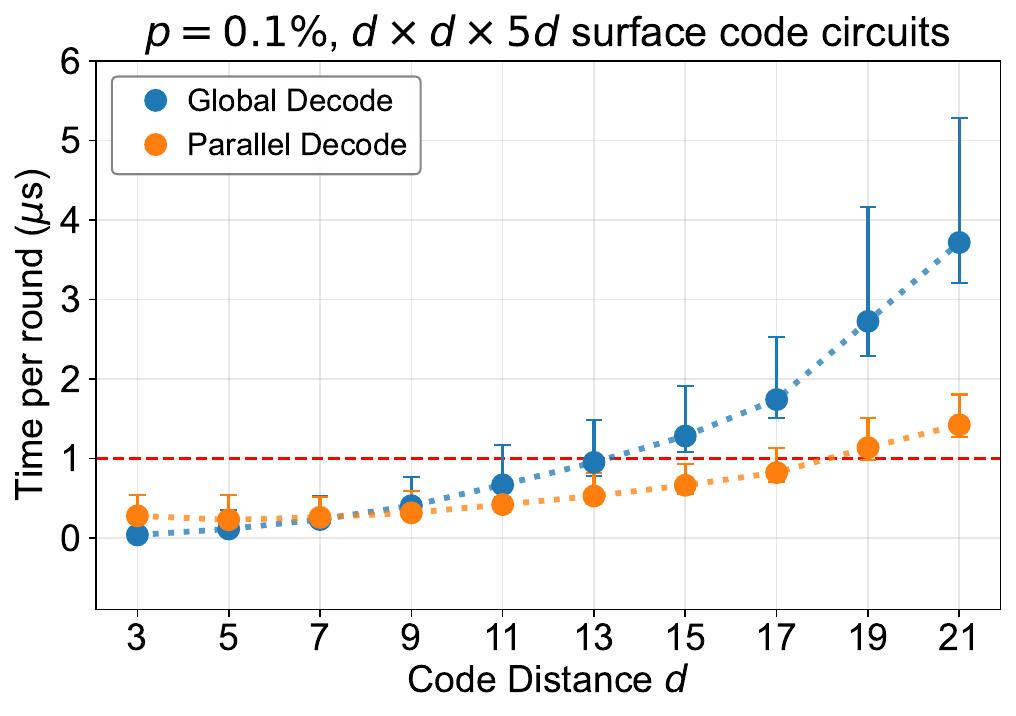}
    \includegraphics[width=0.35\linewidth]{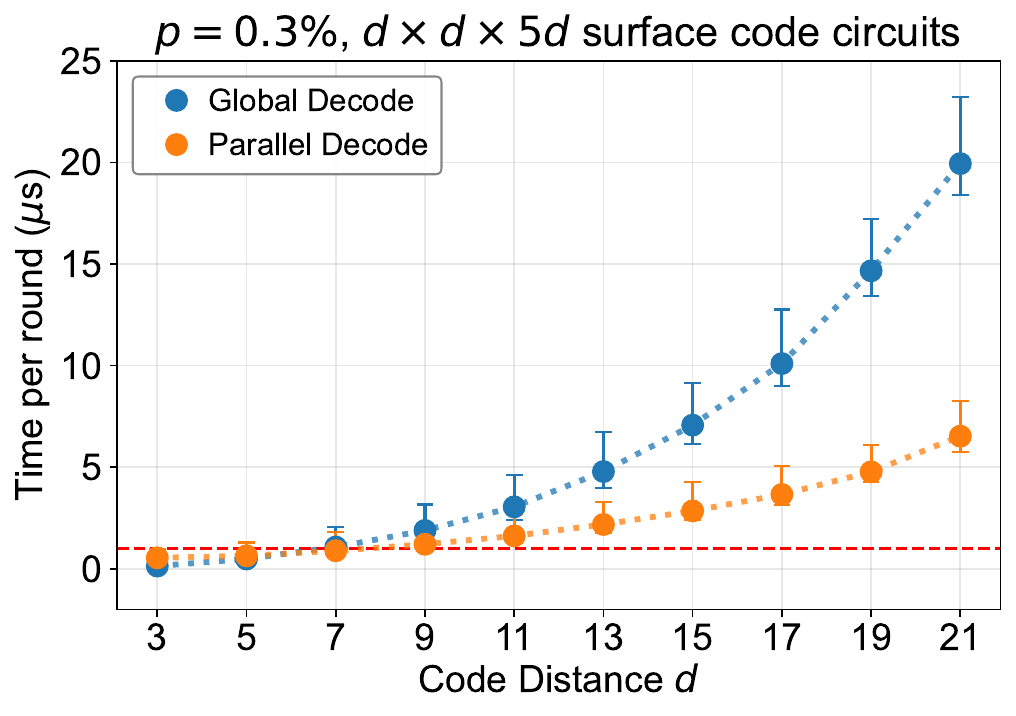}
    \includegraphics[width=0.35\linewidth]{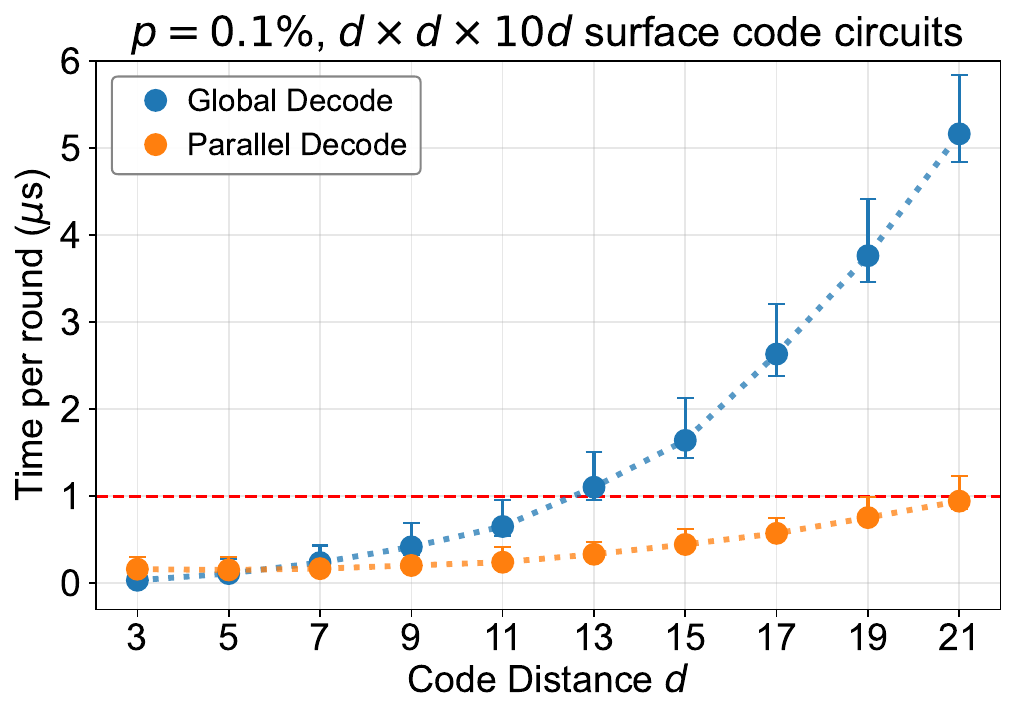}
    \includegraphics[width=0.35\linewidth]{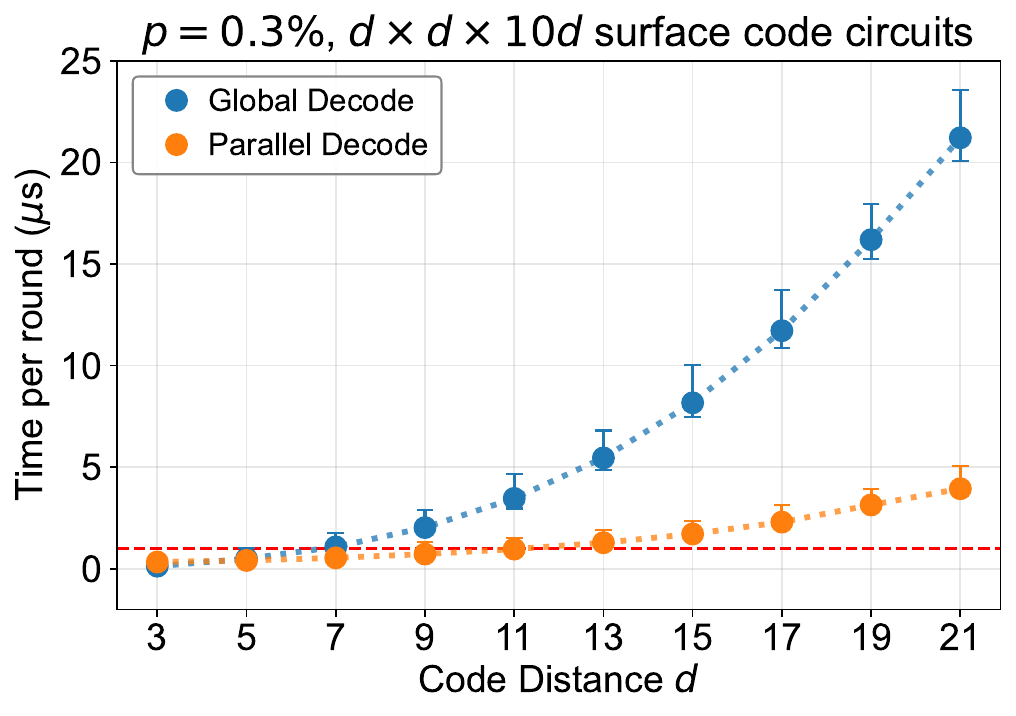}
    \caption{Decoder-level latency for $d \times d \times 5d$ (top) and $d \times d \times 10d$ (bottom) surface-code memory circuits using global MWPM and parallel window with buffer size $b=(d+1)/2$.}
    \label{fig:soft-dl-nd-mwpm}
\end{figure*}

\begin{figure}[tbh!]
    \centering
    \includegraphics[width=0.8\linewidth]{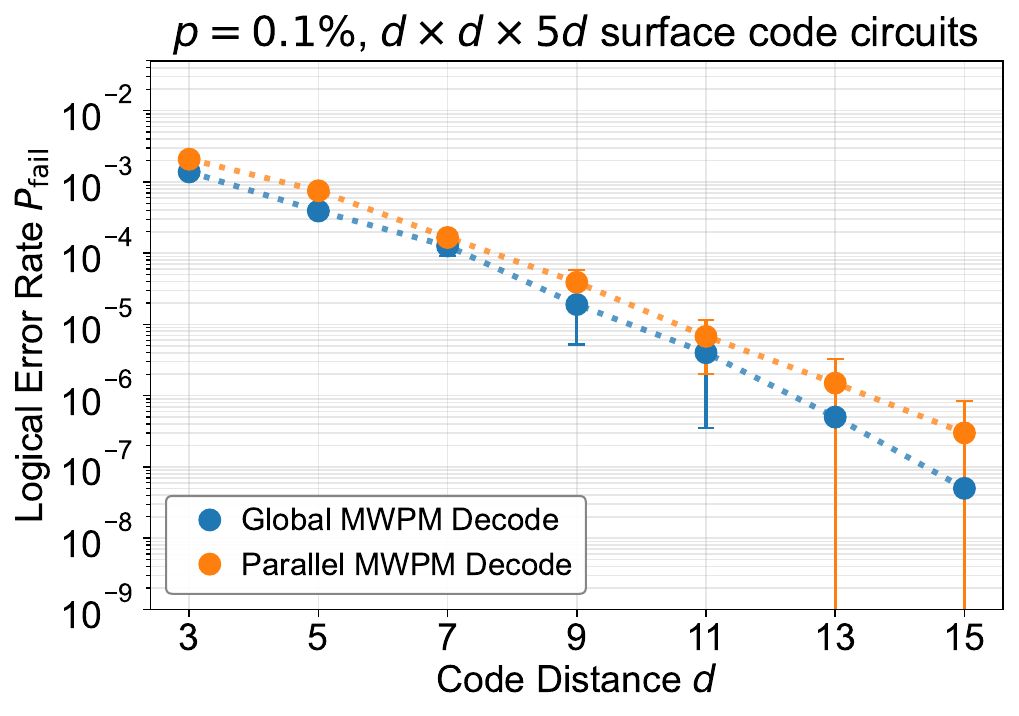}
    \caption{Logical error rate $P_{\mathrm{fail}}$ for $d \times d \times 5d$ surface-code memory circuits at $p=0.1\%$ using global MWPM and parallel window. Sample sizes: $10^5$ shots ($d=3,5,7,9$), $3\times 10^5$ shots ($d=11$), $5\times 10^5$ shots ($d=13$), and $10^6$ shots ($d=15$).}
    \label{fig:parallel-ler-mwpm}
\end{figure}

For fault-tolerant quantum error correction, the decoder will process syndromes over many rounds. Batch (global) decoders handle the full spacetime decoding graph at once, their runtime scales superlinearly with the number of rounds~\cite{dennis2002}. The sliding window decoder partitions the graph into overlapping temporal windows to reduce decoding complexity~\cite{das2022}. These adjacent windows, however, are coupled by a sequential dependency: each window uses the correction output of its predecessor to set its past boundary conditions. The decoding time therefore grows linearly with the number of windows. Eventually, the decoder's processing speed will fall below the syndrome generation rate, resulting in data backlog.

In comparison, the parallel-window decoder~\cite{tan2023,skoric2023}partitions the syndrome stream into independent Type-1 (A) sandwich windows and Type-2 (B) seam windows as shown in Figure \ref{fig:parallel-window}. Each A window consists of a central core region flanked by buffer regions on both sides. The buffer provides contextual information to suppress boundary effects on the core, but only corrections within the core are retained. When a defect lies near the core-buffer boundary, the A window forwards an artificial defect to the adjacent B window so that the seam can be resolved globally.
The B-layer consists of narrow seam windows situated between adjacent A windows. Each B window spans the seam region and receives artificial defects from its two neighbouring A windows. By resolving these seam defects on closed boundaries, the B layer recovers the global topological information that would otherwise be lost to the local A-window partitions.
All A windows are decoded in parallel, followed by all B windows in parallel. 
The decoding time reduces to the sum of the worst-case decoding times for windows A and B, and is independent of the total number of windows if sufficient computing resources are provided.
That is, the decoding throughput scales with the number of available CPU/GPU cores. Both window types use the same or different inner decoder (e.g., minimum-weight perfect matching (MWPM) in our implementation). The parallel-window decoder can achieve comparable accuracy as the global decoder while providing the scalable throughput required for large code distances.

In our demonstration, syndromes are generated round by round. The FPGA streams each new syndrome layer through the TH-Express interconnect to the host decoder and writes it to a rolling buffer. The decoder partitions the decoding graph into windows that progress incrementally as data streams in.
Let $d$ denote the code distance. The step size is $s=(d+1)/2$ and the buffer size is $b=(d+1)/2$, so each A window spans $w=s+2b=3(d+1)/2$ rounds. The number of A windows is $n_A=\lceil t_{dr}/s\rceil$, where $t_{dr}$ is the total number of detector rounds received, and the number of B windows is $n_B=\max(0,n_A-1)$.
The first A window begins decoding once it has received $w$ rounds. Subsequent A windows start at fixed intervals of $s$ additional rounds. 
% All A windows execute concurrently, staggered by $s$ rounds. 
A B window starts after its two neighbouring A windows have completed; because the A windows are launched at fixed intervals, each B window starts with a bounded, deterministic delay. All B windows also execute with an offset timing according to the completion of their parent A windows. After both phases finish, the decoder assembles the final correction set by concatenating the retained matching edges from the cores of all A windows and the edges from all B windows. 
% Each edge specifies two detector endpoints and a fault mask.
% In the Stim DEM framework, these correspond to the estimated physical error mechanisms. 
The decoder maps this edge set to physical qubit indices through a detector-to-qubit lookup table, producing the corresponding corrective Pauli operators. This correction is transmitted back to the FPGA through TH-Express, closing the real-time decoding loop.

We characterize the software-level decoding latency and the logical error rates of the parallel-window decoding. 
Figure~\ref{fig:soft-dl-d-mwpm} shows the software-level decoding latency of global MWPM for $d \times d \times d$ surface-code memory circuits. Each point corresponds to the median of multiple independent syndrome realizations, with error bars indicating the 95\% coverage interval. At $p=0.1\%$, the latency remains below 1~$\mu$s for $d \le 7$ and grows approximately linearly with code distances.
It exceeds the real-time threshold once $d > 11$. Raising the physical error rate to $p=0.3\%$ increases the absolute latency because the syndrome graph becomes denser, but the scaling with $d$ remains similar.

For $d \times d \times 5d$ and $d \times d \times 10d$ memories, parallel-window decoding reduces the median latency compared to global decoding as the distance increases (Fig.~\ref{fig:soft-dl-nd-mwpm}).
The parallel-window strategy achieves around 1~$\mu$s decoding latency at $d=21$ for $10d$ rounds and $p=0.1\%$, while the global decoding achieves around 1~$\mu$s latency at $d=13$. 
% The advantage widens at the higher error rate because denser syndrome graphs amplify the benefit of concurrent window execution.

Figure~\ref{fig:parallel-ler-mwpm} compares the logical error rate $P_{\mathrm{fail}}$ for $d \times d \times 5d$ rounds at $p=0.1\%$ (sample sizes: $10^5$ shots for $d=3,5,7,9$; $3\times10^5$ for $d=11$; $5\times10^5$ for $d=13$; $10^6$ for $d=15$). 
% The accuracy loss of this parallel-window decomposition is modest. 
Parallel window achieves a slightly higher logical error rates than the global decoding. 
% The same trend holds for the ML decoder (Fig.~\ref{fig:parallel-ler-ml}). 
This accuracy loss arises because defects near the core-buffer boundary are resolved locally within each A window, while the B-layer windows recover global topological information on closed boundaries formed by artificial defects forwarded from adjacent A windows. Thus, parallel window trades a slight increase in logical error rate for a substantial latency reduction on large number of syndrome rounds.

\end{document}